\documentclass{emulateapj}

\usepackage{graphics,graphicx}

\begin{document}


\title{The Effect of a Cosmic Ray Precursor in SN 1006?}
\author{Cara E. Rakowski\altaffilmark1, J. Martin Laming\altaffilmark2,
Una Hwang\altaffilmark3, Kristoffer A. Eriksen\altaffilmark4, Parviz
Ghavamian\altaffilmark5, \& John
P. Hughes\altaffilmark4}

\altaffiltext{1}{Space Science Division, Naval Research Laboratory, Code 7671, Washington DC 20375}
\altaffiltext{2}{Space Science Division, Naval Research Laboratory, Code 7674L, Washington DC 20375}
\altaffiltext{3}{Goddard Space Flight Center and Johns Hopkins University}
\altaffiltext{4}{Rutgers, the State University of New Jersey}
\altaffiltext{5}{Space Telescope Science Institute, 3700 San Martin
  Drive, Baltimore, MD 21218} 




\begin{abstract}

Like many young supernova remnants, SN 1006 exhibits what appear to be
clumps of ejecta close to or protruding beyond the main blast wave. In
this paper we examine 3 such protrusions along the east rim.
They are semi-aligned with ejecta fingers behind the
shock-front, and exhibit emission lines from O VII and O VIII.
We first interpret them in the context of an upstream medium modified
by the saturated nonresonant Bell instability which enhances the growth of
Rayleigh-Taylor instabilities when advected postshock.
We discuss their apparent periodicity if the spacing is determined
by properties of the remnant or by a preferred size scale in the
cosmic ray precursor.
We also briefly discuss the alternative that these structures have an origin in the ejecta
structure of the explosion itself. In this case the young evolutionary age of
SN 1006 would imply density structure within the outermost layers of the
explosion with potentially important implications for deflagration and
detonation in thermonuclear supernova explosion models.

\end{abstract}

\keywords{acceleration of particles --- cosmic rays --- ISM: supernova remnants --- shock waves --- supernovae:
  individual(SN 1006)}

\section{Introduction}
Supernova remnants (SNRs) have long been suspected as the sites of
cosmic ray (CR) diffusive shock acceleration, where $\sim$10\% of the
supernova energy would be required to be transmitted to CRs to satisfy
observations. Partial confirmation of this idea came with ASCA
observations of SN 1006 \citep{koyama95,reynolds96}, which revealed NE
and SW limbs dominated by X-ray synchrotron radiation from CR
electrons. Direct evidence of CR ions, presumably the dominant
CR component, has been harder to find, for the most part being
limited to the observation of $\gamma$-rays resulting from the
reaction $p+p\rightarrow \pi ^0 + X$ followed by $\pi ^0\rightarrow
2\gamma$ decay. \citet{eriksen11} interpreted newly discovered, regularly
spaced stripes of non-thermal emission in Tycho's SNR as regions where
the pre-shock medium has been significantly modified by
$\sim10^{15}$eV ions.
Indirect evidence for CR ions comes from the
separation between the forward shock and the contact discontinuity in
young SNRs \citep{warren05,cassam08,ferrand10}. For example, in SN
1006, the contact discontinuity is predicted to be less than
0.86 of the forward shock radius, but is observed to be larger than
that around almost all of the remnant \citep{cassam08,miceli09}. This has
been interpreted in SN 1006 and Tycho's SNR as being due to energy losses to CRs at the
forward shock \citep{warren05,cassam07,cassam08,miceli09}, which leads to a
greater shock compression ratio and a thinner shell of shocked CSM
\citep{blondinellison01,fraschetti10}.
However, only in cases of extreme CR energy losses can a fraction of the ejecta in Rayleigh-Taylor (RT)
instability approach or overtake the forward shock.

Density inhomogeneities in either the interstellar medium \citep{jun96a} or
the ejecta \citep{wang01,blondin01} can also enhance the growth of RT
instabilities potentially allowing them to reach the forward shock.
Originally motivated to explain the existence
of radial magnetic fields in young SNRs \citep{jun96b}, \citet{jun96a}
explored the effect of turbulence induced as the SNR
forward shock encounters density clumps (of a factor of 5) in the
preshock medium.
The vorticity induced by the interaction of the forward shock
with the overdense clumps is advected postshock. Upon reaching the contact
discontinuity, the vortices transfer their rotational energy to the RT
fingers. This enhances the growth of the fingers such
that ejecta clumps may protrude beyond the blast wave.
In some specific cases it was found that twenty times larger density inhomogeneities
would be required in the ejecta to achieve the same result
\citep{wang01,blondin01}.

Upstream density clumping may be intrinsic to the pre-shock medium, or
induced by a CR shock precursor. In the scenario of nonresonant
magnetic field amplification \citep{bell04,bell05},
density clumping by a factor of $\sim$3 is
produced upstream of the shock as the magnetic field amplification
saturates. Magnetic field amplification, also taken by some
as indirect evidence for proton acceleration at shocks, is suggested by the
narrowness of rims of synchrotron radiation at SNR shocks
\citep{vink03,long03,yamazaki04,warren05,cassam07,volk05}. Modeling
the widths allows one to estimate the magnetic field to be in the
range 100 - 500 $\mu$G, far larger than can be attained by the typical compression of
ambient magnetic field. Fields of this magnitude may speed up the CR
acceleration process, allowing SNRs to generate CRs up to and
beyond the ``knee'' in the CR spectrum, at about
$10^{15}-10^{16}$ eV.

Of all known SNRs, SN 1006 offers the most promising test of the action of CRs to
produce density clumping ahead of the forward shock. It is the likely remnant of a Type Ia supernova, located
500 pc above the galactic plane and expanding into low density
interstellar medium.  A H.E.S.S. detection of
SN 1006 has been reported \citep{naumanngodo08}, consistent with
an ambient gas density of 0.05 cm$^{-3}$ \citep{acero07}.
Proper
motions have been measured in the NW in the optical, 0.28'' yr$^{-1}$
\citep{winkler03} corresponding, at a distance of 2.2 kpc, with a shock
velocity of 2900 km s$^{-1}$\citep{ghavamian02}.
In the NE, 0.48'' yr$^{-1}$ expansion was measured in X-rays \citep{katsuda09}
giving 5000 km s$^{-1}$, indicating higher density in the NW than elsewhere.
Hamilton et al. (1997; 2007) derive a reverse shock velocity of 2700 km s$^{-1}$, and determine the expansion velocity of ejecta entering the reverse shock to be 7000 km s$^{-1}$, from HST observations.

In this paper, we examine Chandra images and discuss structures seen
to protrude beyond the eastern portion of the shock front.
Ejecta clumps with similar periodicities are
seen behind or near the blast wave elsewhere in SN 1006, most
prominently around the southern rim.
%
We chose to
study these three features because they extend past the main shock and
have deep enough observations for minimal spectral analysis. A more
definitive study of all periodic structures in SN1006 is defered to
future work, thus numerical
results reported herein may not be applicable for all of SN1006.

\begin{figure}[bh]
\includegraphics[width=3.2in]{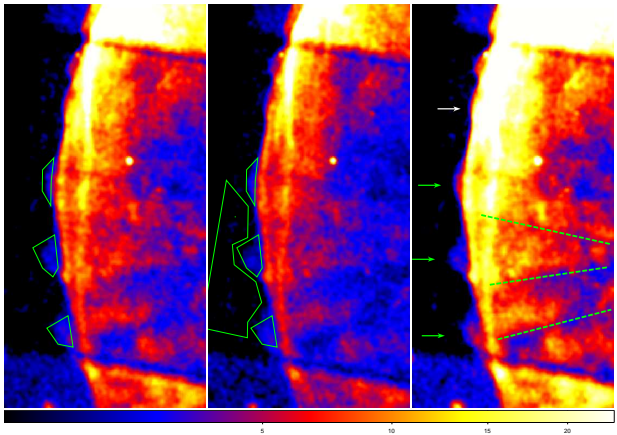}
\caption{From left to right, the 2000 and 2008 Chandra observations of the eastern portion of
SN1006 from 0.2 to 5.0 keV and a merged image where the 2008 image was
shifted by the proper motion measured by \citet{katsuda09}. All three panels were
binned by 4 pixels, smoothed and use square root scaling to show both the faint and brighter
features. The overlaid regions show where spectra were extracted
from the 2008 observation.}
\vspace{-0.14in}
\end{figure}

\section{Observations}

The eastern portion of SN1006 has been imaged multiple times with
Chandra and XMM, and the features under discussion here are evident in
all observations. However for the sake of higher statistics we focus
only on the two deepest Chandra observations in 2000 July 10, (Obsid
732) and 2008 June 24 (Obsid 9107). As in
\citet{katsuda09} we exclude times of high background from the 2000
observation but found no flares in the 2008 observation. To further
reduce the background we consider only the band in which the emission
from SN1006 dominates, from 0.2 to 5.0 keV for imaging, and 0.3 to 6.0
keV for spectral analysis.

Figure 1 shows the region of interest in each long observation
and a merged image where the 2008 observation was shifted 3\farcs85 along a line 7.5$\degr$ S of W to match the proper motion found in
\citet{katsuda09}. The features outside of
the main shock are highlighted with green polygons that mark the
spectral extraction for the 2008
observation. (The 2000 spectra were extracted from proper-motion-shifted regions).
The southernmost feature extends below the spectral
extraction region, intersecting the bright shock below the chip gap.
The coordinates of the polygons are the same in both frames, such that
one can clearly see the movement of the outer limits of the features
from 2000 to 2008. The proper motion of the features appears
comparable to the main bright shock, but is difficult to determine exactly
given their low surface brightness. A fourth protrusion to the north
that was too narrow to extract a spectrum from
is marked in the third panel with a white arrow. Similar features can
be seen further north.
They do not extend so far in advance of the forward shock, and have
no obvious connection with structures extending post shock.

\begin{deluxetable}{lcc}
\tablecaption{Size scales of upstream features\label{tbl-1}}
\tablewidth{0pt}
\tablehead{
\colhead{} & \colhead{angular} & \colhead{linear} \\
\colhead{} & \colhead{(arcseconds)} & \colhead{($10^{18}$ cm)$^{a}$}}
\startdata
azimuthal length & 83, 41, 46$^{b}$ & 2.7, 1.5, 1.4\\
radial extent & 25, 20, 10 & 0.8, 0.7, 0.3 \\
spacing$^{c}$ & 145, 125, 125 & 4.8, 4.1, 4.1 \\
\enddata
\tablenotetext{a}{Assuming 2.2 kpc distance to SN 1006}
\tablenotetext{b}{Three features, from south to north}
\tablenotetext{c}{Measured from ``leading edge'' to leading edge of
 each feature as opposed to the size of the darker regions between
 features. The northernmost size is measured from the feature
marked with a white arrow in Figure 1.}
\end{deluxetable}

In Table 1 we present the size scales of the upstream features.
The spacing between the three features
corresponds to a wavelength of $\sim 4\times 10^{18}$ cm at the
distance of SN 1006 and assuming all are in the plane of the sky.
Looking to the remnant interior one can see clumpy extended fingers of
brighter emission with a similar spacing
underlined with dashes in Figure 1. The most obvious of these extends
from behind the southernmost feature, but others are suggested
behind each feature.
The peak
brightness of the protrusions is between 15\% and 45\% of the peak
brightness of the main shock front immediately behind it.



\begin{figure}[h]
\includegraphics[width=2.8in]{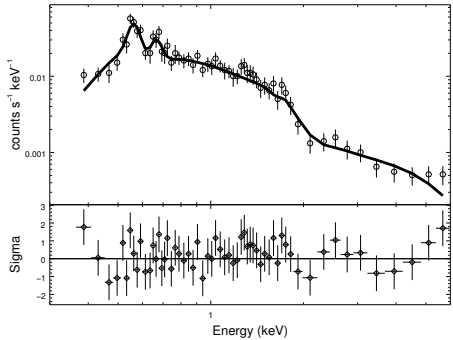}
\caption{
The combined spectrum of all three regions ``ahead'' of the shock from
the 2000 observation, fitted with an absorbed power-law model plus two
Gaussian lines for the H-like and He-like O.
 }
\end{figure}

Figure 2 shows the combined spectrum of all three features ahead of
the shock in the 2000 data. A background region from the same
chip that excluded the features was chosen and modeled rather than
subtracted and the spectra were binned to allow the use of $\chi ^2$
statistics. The total counts in the 2000 data between 0.3 and 6.0 keV
were 1650 and 1033 for the 3 features and the
background region respectively. Table 2 summarizes the best fit values and
3$\sigma$ confidence intervals for a simultaneous fit of the 2000 and
2008 data.
An absorbed power-law model with two additional Gaussian
lines fits the data well, with an appropriate $N_{H}$ for SN 1006, and
a power-law index consistent with fits to a nearby region of the
bright shock. The addition of two Gaussian lines for He-like and
H-like O is supported by an Ftest at the 99\% significance level.
%
%
%
%
Alternatively, following the SE model of \citet{miceli09},
the data are fit equally well with a synchrotron and non-equilibrium
ionization thermal component. In this fit we allowed only the
normalizations, powerlaw index $\gamma$, temperature $kT$, and
ionization timescale $n_{e}t$ to vary. Their analysis yielded a super-solar
oxygen abundance which matches our spectrum well, the only difference
being the best-fit ionization timescale for the 3 features is
nominally lower than that of the shocked region as a whole (at
1$\sigma$, but not at 3$\sigma$).

\begin{deluxetable}{lcc}
\tablecaption{Spectral model fits for the J2000 and J2008 data\label{tbl-2}}
\tablewidth{0pt}
\tablehead{
\colhead{Parameter} & \colhead{Powerlaw $+$
Lines} & \colhead{Miceli SE, Powerlaw $+$ NEI }}
\startdata
$N_{H}$ (cm$^{-2}$) & $5.8^{+16}\times 10^{20}$ & $7\times 10^{20}$  \\
$\gamma$ & $2.7^{+0.6}_{-0.4}$ & $2.7^{+0.3}_{-0.5}$\\
Line (eV) &  $561^{+10}_{-11}$ & \\
Line (eV) &  $669^{+21}_{-20}$ & \\
$kT$ (keV) & & $0.57^{+0.6}$\\
$n_{e}t$ (cm$^{-3}$ s) &  & $5.7^{+24}_{-3.1}\times 10^{8}$ \\
$\chi^{2}$ (dof) & 200.5(197) & 210 (199) \\
\enddata
\end{deluxetable}

\section{Discussion}
\subsection{Overview}
The most relevant observed characteristics of the three pre-shock
features are the generic size-scale of $\sim 4\times 10^{18}$cm, a
continuum that is consistent with the synchrotron emission from the
bright shock, and the presence of He-like and H-like O lines that
imply an ionization age of $\sim 5.7^{+24}_{-3.1}\times 10^8$ cm$^{-3}$ s.
This ionization age, comparable to the remnant lifetime given the ambient density of SN 1006,
argues against an interpretation of the features
as the denser portions of a CR precursor. The likely
supersolar abundances with a synchrotron dominated continuum, relative brightness of the preshock features,
and small radius of curvature also support the idea that these are
ejecta fingers that have breached the shock front.

Hereafter, we discuss possible origins for the spacing of the
pre-shock features. (1) Anomalous viscosity determines the preferred
RT wavelength. (2) The saturated state of nonresonant magnetic field
amplification creates cavities and higher density regions at a
particular scale. (3) The ejecta fingers originate in the explosion itself.

\subsection{Ejecta Fingers Spaced by Anomalous Viscosity?}
If the preshock features seen are the extension of the ejecta fingers
themselves \citep{cassam08} then the spacing between them may reflect the
wavelength of the R-T instability. In the absence of a periodic perturbation,
this depends on the kinematic viscosity (the product of the sound speed and largest eddy size: $\nu = c_{sound}l_{eddy}$).

In stationary incompressible media the Rayleigh-Taylor linear growth
rate is approximately $\Gamma = \sqrt{\nu ^2k^4+gk}-\nu k^2$  which has
a maximum at mode wavenumber $k=\left(g/\nu ^2\right)^{1/3}/2$ \citep{plesset74},
where
$g\simeq 0.0049$ cm s$^{-2}$ for SN 1006 is the deceleration of the
plasma \citep[estimated from the SNR dynamics using][]{truelove99}, and the ejecta density is assumed
much greater than that of the shocked ISM.
Exponential growth goes over to power law growth as displayed in \citet{blondinellison01} when expansion is included. Taking the observed
wavelength of $4\times 10^{18}$ cm, we find $\nu\sim 10^{25}$
cm$^2$s$^{-1}$ and a maximum linear growth rate $g^{2/3}/2\nu
^{1/3}\sim 6\times 10^{-11}$ s$^{-1}$. This is slow, allowing only 2 e-folding
times during the 1000 yr age of SN 1006. Additional vorticity is probably needed to speed up the process.

We can
compare this to an estimate of the viscosity from the product of the
sound speed in the shocked interstellar medium (about $4\times 10^8$
cm s$^{-1}$) and a length scale determined by the separation between
the forward shock and contact discontinuity (about 1 pc) to find
$\nu\sim 10^{27}$ cm$^2$s$^{-1}$.
Thus either the viscosity is anomalous by a factor of $\sim 10^{-2}$, or the wavelength of the instability is
determined by different physics, i.e. the CRs.

\subsection{Clump Spacing Determined by Cosmic Rays?}



We explore the case that the ejecta finger spacing is determined by structure in the CR precursor
where non-resonant magnetic field amplification has reached saturation
\citep{bell04,bell05}.
Structure arises because the highest
energy CRs, which remain unmagnetized 
inhabit cavities in the
upstream medium which are evacuated by the amplified magnetic
field. The plasma and magnetized CRs are swept into
cylindrical ``walls'' surrounding the cavities where the originally
quasi-parallel shock has now become quasi-perpendicular. Interpreting
the observed structures %
as ejecta fingers protruding into the cavities,
the cavity
radius, observed here to be around 55'', or $2\times 10^{18}$ cm at
the 2.2 kpc distance of SN 1006 is similar to the unmagnetized CR
gyroradius $r_{g}$, so
\begin{equation}
{\gamma mc^2\over qB} \sim 2\times 10^{18}{\rm cm}\Rightarrow B\sim 2\times 10^{-12}\gamma {\rm G},
\end{equation}
where $\gamma$ is the Lorentz factor for a high energy unmagnetized
CR and $B$ is the amplified magnetic field. The CR
rest mass and charge are $m$ and $q$, and $c$ is the speed of light.
To estimate $B$ and $\gamma $, we consider a condition for the existence of the nonresonant instability \citep{bell04}
\begin{equation}
{qB\over\gamma mc^2} < k_{\Vert} < {J_{CR}B\over\rho cv_A^2}={4\pi n_{CR}qv_s\over\gamma cB}\Rightarrow B^2 < 4\pi mv_scn_{CR}
\end{equation}
where $v_A^2=B^2/4\pi\rho$ and $J_{CR}=n_{CR}qv_s/\gamma$ is the CR current in terms of the total number density of CRs $n_{CR}$, (of which the
unmagnetized fraction $n_{CR}^{\prime}=n_{CR}/\gamma$), the charge $q$, and the shock
velocity $v_s$ at which CRs are assumed to stream.\footnote{We are taking the CR distribution function $f_{CR}=n_{CR}p_{inj}/4\pi p^4$, $n_{CR}=\int _{p_{inj}}^{p_{max}}f_{CR}4\pi p^2dp$ with $p_{max} >> p_{inj}$, and $n_{CR}^{\prime}=\int _{p_{min}}^{p_{max}}f_{CR}4\pi p^2dp\simeq n_{CR}p_{inj}/p_{min}\simeq n_{CR}/\gamma $. The injection momentum is
$p_{inj}\sim mc$, so that $\gamma _{inj}\sim 1$, and $p_{min}\simeq\gamma mc^2$ is the minimum momentum of the unmagnetized CRs.} For SN 1006
with $v_s\simeq 5000$ km s$^{-1}$, $B^2 < 3\times 10^{-4}n_{CR}$. The
total CR number density in terms of $\eta =P_{CR}/\rho v_s^2$, the ratio of CR pressure to shock ram pressure, with
$P_{CR}=E_{CR}/3=\int _{p_{inj}}^{p_{max}}f_{CR}pc 4\pi p^2dp/3$,
is
\begin{equation}
n_{CR}={3\eta n_iv_s^2\over\ln\gamma c^2}
\end{equation}
where $n_i\simeq 0.05$ cm$^{-3}$ is the ion density in the upstream
medium. Hence $B^2\sim 10^{-9}\eta$, or with $\eta = 0.1$, $B\sim 10
\mu$G then $\gamma \sim 6\times 10^6$.
At a position further to
the north, \citet{morlino10} infer a value of $\eta\simeq 0.29$, which
results in higher values of $B\sim
17\mu$G and $\gamma\sim 10^7$. Both these values imply
$E=10^{15}-10^{16}$eV, which is past the ``knee'' in the CR
spectrum.

\subsection{Ejecta Fingers Associated with SN Explosion?}

The possibility remains that the ejecta structure interpreted as due
to RT instability above is in fact related to the Type Ia
explosion, as in e.g. the O-DDT model of \citet{maeda10a} illustrated in
\citet{maeda10b}.
In other SN Ia remnants, such as Tycho, similar clumps to the ones considered
here show the clear presence of Fe ejecta and as a result have been interpreted in this way. Delayed detonation models \citep[e.g][]{gamezo04} appear to
be the most plausible explosion model for a Type Ia SN \citep[see e.g.]{badenes06}.
Ashes burnt by the deflagration may exhibit structure
\citep{maeda10a,jordan08}, whereas large density gradients appear to be absent from ejecta burnt by a detonation wave \citep[e.g][]{maeda10a,meakin09}.
Unburnt ejecta {\em exterior} to these regions
may survive due to expansion ahead of the deflagration front, with the density diminishing sufficiently
to allow them also to escape burning in any subsequent detonation as
well.

In SN 1006, both spectrally and dynamically, it appears that
the reverse shock has only recently encountered
regions that were subject to thermonuclear burning.
Suzaku recently detected Fe K emission with a low
ionization age consistent with the Fe ejecta being more recently
shocked than the other ejecta (Yamaguchi et al. 2008).
The ejecta expansion velocity at the reverse shock in the observations
of Hamilton et al. (1997; 2007) is close to the predicted
outer extent of Fe produced in the C-DDT and C-DEF models of Maeda et al. (2010a), and also that in the various models tabulated by Badenes et al. (2005). Hamilton et al. (1997) comment that the ISM density at the back of the SNR is likely to be lower than elsewhere, and so the reverse shock in other regions of the remnant may have progressed further into the ejecta, shocking some of the Fe as is now seen by
Suzaku (Yamaguchi et al. 2008).
By comparison, in Tycho's SNR, Badenes et
al. (2006) find that the reverse shock has reached all but the last 0.4
$M_{Sun}$ of ejecta. Thus these Ia SNRs are at significantly different evolutionary stages in terms of the location of the reverse shock in the ejecta.
An ejecta origin for the protrusions in SN
1006 would have different implications for the locus of deflagration
and detonation burning than for similar clumps in Tycho, in terms of the location
of deflagration ashes and whether their structure can survive a
detonation wave, as in the asymmetrical O-DDT model of Maeda et al. (2010a).

\section{Conclusions}



The structures observed exterior to the blast wave along the eastern
limb of SN 1006 are
likely to be metal-rich ejecta based on their probable supersolar abundances of O.
We believe that a mechanism combining the ideas of \citet{jun96a} and
\citet{bell04} is one way to bring the ejecta so close to the blast
wave. Upstream density inhomogeneities of only a
factor of 5 in the ISM could allow some RT fingers to overtake the
blast wave, with the cavities and density enhancements that are a
natural consequence of CR magnetic field amplification.

We contend that prior explanations for ejecta overtaking the shock are
insufficient if acting alone.
Invoking enhanced shock compression following CR
losses, \citet{blondinellison01}, needed post-shock compressions by a
factor of 21
to obtain RT fingers that actually overtook the forward shock. This
would imply implausibly high energy losses \citep[77\% of shock ram
pressure lost as CRs, from equation 2 of][]{liang00}. The
recent more conservative simulations of \citet{fraschetti10} do not
show RT fingers that exceed the blast wave radius, but do show that
2\% of the ejecta material reaches within 95\% of the forward shock
radius \citep[see also][]{wang11}.
%

If the protrusions are RT fingers, we can interpret the size scale of the pre-shock features in terms of properties
of the SNR in this sector of the shell. If
the size scale is determined by
the properties of the SNR plasma, assuming that the CR precursor
itself establishes no preferred wavenumber for RT growth, then the
viscosity would need to be anomalous by a factor of $\sim
10^{-2}$. If instead the CR precursor develops a preferred size scale \citep{bell04,bell05}, and this determines the wavelength of RT growth, the spacing may be interpreted in terms of CR parameters to give
$B\sim 10 \mu$G and $\gamma \sim 6\times 10^6$. This implies CR
acceleration up to the knee in SN 1006.

The other tantalizing possibility is that these structures have an
origin in the ejecta. If deeper observations revealed Fe as is seen in
some ejecta clumps in Tycho's SNR, the clumpiness, Fe, and
low evolutionary age of SN 1006 would imply that
deflagration ashes survived near the exterior of the exploding white
dwarf.




CER and JML acknowledge support from NASA contract NNH10A009I, and basic
research funds from the Office of Naval Research. KAE and JPH
acknowledge support from Chandra grant number GO9-0078X to Rutgers
University. PG acknowledges HST grant GO-11184.07A at STScI.


\end{document}